\documentclass[twocolumn, superscriptaddress, 11pt]{revtex4}

\usepackage{amsmath,amssymb,graphicx}

\begin{document}

\title{Gravitational wave emission from quadrupole Josephson junction device}

\author{Victor Atanasov}
\affiliation{Faculty of Physics, Sofia University, 5 blvd J. Bourchier, Sofia 1164, Bulgaria}
\email{vatanaso@phys.uni-sofia.bg}

\begin{abstract} 
We suggest a hybrid quantum mechanical/classical set up capable of gravitational wave emission. The proposed device consists of two superconducting tunneling junctions which act as quantum voltage-to-frequency converters and produce oscillation of the charges (masses) in the superconducting condensate. The classical interpretation of the set up is associated with the particular arrangement which leads to suppression of the dipole radiation and formation of time dependent quadrupole mass moment - the gravitational wave source.   The proposed device converts electrical energy/momentum into gravitational ones thus suggesting a form of propulsion. Provided the reverse effect: curvature-to-voltage conversion is possible, the device represents an emitter-receiver pair for gravitational wave based communication (favorable $1/r$ fall off with distance).     
\end{abstract}

\maketitle

Gravitational wave emission is associated with the conversion of the kinetic energy of a system of masses (time dependent quadrupole mass moment) into geometrical energy\cite{LL}. The mechanism is purely classical and is contained in Einstein's field equations. 

Interestingly, when one considers the effects of curved space-time on a quantum mechanical system, they are conveyed via geometric potential that springs out of the kinetic energy operator. From a quantum mechanical point of view the curvature induced potential is very similar to the kinetic energy $E_{kin}$ of a quantum system\cite{podolsky}
\begin{equation*}
E_{kin} \propto \quad \frac{\hbar^2}{m} \nabla^2,
\end{equation*}
where $\hbar$ is Planck's constant and  $m$ is the mass of the quantum particle/system. Note $\nabla^2$ is inversely proportional to the square of a distance, that is curvature.

The curvature induced potential emerges when a quantum system is constrained to abide  curved space-time. In this case the quantum system is no longer free.  DeWitt found that the Hamiltonian of the quantum system in curved space exhibits an extra term $V_{geom}$ containing the Ricci scalar curvature (inversely proportional to the square of a distance)\cite{dewitt}: 
\begin{equation*}
V_{geom} \propto \frac{\hbar^2}{m} R. 
\end{equation*}
The only ambiguity in his Feynman-inspired approach being an undetermined numerical factor.  

Again, the curvature-induced quantum interaction is compulsively resemblant to a kinetic term\cite{jj}. The kinetic energy-like origin of the curvature-induced quantum interaction was confirmed in the thin layer confining potential approach\cite{JK, daCosta}. Therefore, classically, as well as quantum mechanically, the curvature of space-time and the kinetic energy of an  embedded astrophysical or condensed matter quantum systems are related.

\begin{figure}[b]
\includegraphics[scale=0.60]{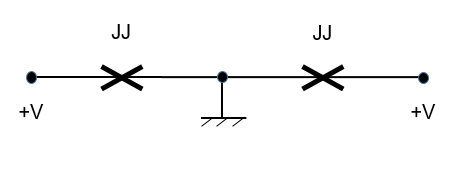}
\caption{\label{fig1} A quadrupole superconducting Josephson junction device.}
\end{figure}

In this paper we explore the suggested relation and propose a quantum mechanical/classical hybrid superconducting device which is a promising candidate for a controlled kinetic into gravitational wave's energy conversion. The generation of artificially induced gravitational waves via condensed matter systems represents an unique opportunity to explore gravitational physics beyond its observational stature and is worth pursuing due to present day technological capabilities when it comes to device manufacture.

In a condensed matter system the charge distribution entails a mass distribution as well. The mass in the condensed matter context is concentrated in the positively charged
lattice of ions, whereas the electronic subsystem is three to four orders of magnitude lighter. However, the electronic subsystem (especially the part which is devoted to conductivity, transport, many-body and tunneling phenomena) is extraordinary mobile. The maximum characteristic frequencies of the collective excitations in the periodic and elastic arrangement of atoms in condensed matter are determined by the phonon dispersion. In the case of niobium, phonon frequencies can go well in the THz region\cite{Nb} but restricted in amplitude (energy of the high-frequency quasi-particle). Highly energetic processes in the lattice are generally classical and very slow. However, the electronic subsytem is as capable of handling oscillations well in and beyond the UV region (Bremsstrahlung radiation is in the X-ray region), as is capable of tolerating large energy excitations. Electronic processes can be so fast that during the process the lattice experiences hardly any displacement, therefore can be assumed static. 

When we consider a fast electronic process which entails a charge multipole moment, we would assume a non-zero mass multipole moment as well. The relation that joins these two distributions in the condensed matter context has the form $\rho=k \rho_{q},$ where $\rho /\rho_{q}$ are the mass/charge density: [kg/m$^3$] vs. [C/m$^3$]. The coupling constant has the dimension of [kg/C] and is typically very small and of the order of
\begin{eqnarray}
k \propto \frac{m_{e}}{e}=5.6 \times 10^{-12} \quad {\rm kg \, C^{\;-1}}.
\end{eqnarray} 

Since in condensed matter systems one can control the charge distribution, we can expect an artificially created mass distribution as well provided the electronic system is violently jerked while the positively charged nuclei remain in their fixed lattice positions. Obviously, the scale of the mass distribution is of the order of the coupling constant as compared to the charge distribution, that is extremely small. The objective of the paper is to suggest a process for which the small mass that undergoes an oscillation can be largely leveraged by the high frequency of the oscillation.

Next, the main requirement for gravitational wave emission is the time dependent mass quadrupole moment tensor (traceless $D_{\alpha \alpha} =0$) defined by $
D_{\alpha \beta} = \int \rho \left( 3 x_{\alpha} x_{\beta} - r^2 \delta_{\alpha \beta}  \right) dV.$
 Suppose an axial symmetry (with respect to z-axis), then the tensor can be cased into diagonal form where $D_{xx} = D_{yy} = - D_{zz}/2.$ Moreover, when considering motion along the z-axis alone (we can set $x=0$ and $y=0$):
$ D_{zz} = 2 \int \rho z^2 dV.$

As a result, $ \dddot{D}_{\alpha \beta}^{\quad 2} = 3\dddot{D}_{zz}^{\; 2}/2
$ and the power of gravitational wave emission equals
\begin{eqnarray}
- \frac{d E}{dt} &=& \frac{G}{45 c^5}  \dddot{D}_{\alpha \beta}^{\quad 2}
\end{eqnarray} 

Now, suppose one constructs an axially symmetric superconducting device depicted on Fig.\ref{fig1}. It consists of two superconducting Josephson junctions sharing a common ground point which splits the device. Provided a constant DC voltage $V$ is applied at the opposing ends (in the absence of external electromagnetic fields), then the Josephson junctions will generate an AC current which frequency is voltage dependent $
I(t)=I_c \sin{(\omega t)},$ 
where 
\begin{eqnarray}
\omega = \frac{2eV}{\hbar} \sim 3.2 \times 10^{15}  \quad{\rm rad \; s^{-1}\; V^{-1} }
\end{eqnarray} 
Therefore, connecting the junction to a voltage source immediately produces alternating currents with enormous frequency $\sim 10^{15}$ Hz/V (no net current unless put in the AC regime so that the time average is non vanishing). These alternating currents set the law of motion of charges as an oscillation around their equilibrium positions with frequency $\omega$ and some unknown amplitude $A$
\begin{eqnarray}
z(t)=A\sin{\omega t}.
\end{eqnarray} 
 
 Due to the particular configuration of the two junctions in the device it realises ''colliding'' trajectories of the particles, vanishing dipole moment and non vanishing quadrupole one:
\begin{eqnarray}
\dddot{D}_{zz}^{\; 2}= 2^7 A^4 \omega^6  k Q \sin^2 {(2 \omega t)},
\end{eqnarray} 
where $Q=\int \rho_{charge} dV$ is the total charge that is oscillating, which can be assumed as the charge due to the superconducting pairs. The geometric configuration of the two junction in tandem with the choice of the symmetric electric power feed represent the classical aspect of this otherwise (due to the superconducting tunneling junctions) quantum device.

The vanishing electric dipole moment is a product of the symmetry of the system of oscillating charges which leads to the creation of equal but oppositely directed dipole moments (and accelerations) which cancel the electromagnetic dipole radiation in the far field region. There is no, a priori, necessity for the cancellation of the electric dipole radiation for the gravitational wave emission to take place. However, since we are not aware of the thermodynamic (statistical) probability of gravitational wave emission when it competes with electromagnetic wave emission, we presume that the system would relax its energy with greater probability via the gravitational process when the main electromagnetic one is suppressed.

The time averaged 
$\langle \dddot{D}_{zz}^{\; 2} \rangle_t= 2^6 A^4 \omega^6  k Q$ is non-vanishing
and therefore the mean power of gravitational wave emission turns out to be
\begin{eqnarray}
P_{GW} &=& \frac{G}{45 c^5}  2^6 A^4 \omega^6  M_{pairs}\\
\nonumber &\approx& 4\times 10^{-53} A^4 \omega^6  M_{pairs}
\end{eqnarray}  
 where $M_{pairs}=kQ$ is the mass of the superconducting condensate (not the condensed matter system). Suppose that the amplitude of oscillation $A \sim \mu{\rm m}$ and $M_{pairs} \sim {\rm ng},$ while the DC voltage is at 1V, then 
\begin{eqnarray}
P_{GW} &\approx& 40 {\rm W}
\end{eqnarray}  

While the gravitational wave emission will complete with electromagnetic wave emission, it will certainly not exceed the mean Josephson junction energy
\begin{eqnarray}
E_{T} &=& \frac{\hbar}{2e}I_c
\end{eqnarray}  
and the mean power of the emission would not exceed
\begin{eqnarray}
P_{J}=\frac{E_{J}}{T} &=& \frac{\pi \hbar}{e} \frac{I_c}{\omega} \approx 2 \times 10^{-15} \frac{I_c}{\omega}.
\end{eqnarray}  
In the discussed unit voltage drop across the junction case and for a Nb/AlOx/Nb liquid-He operated 1 cm$^2$ Josephson junctions, where $I_c \sim kA,$  $P_{J} \approx {\rm kW}.$ This is an additional assertion that such a device can in theory produce powerful enough gravitational wave emission (provided detectors for such high frequency gravitational waves existed).

An important issue remains to be duscussed if the proposed effect is to be physically relevant or not, namely, the Josephson current oscillations might not imply
mass oscillations, since the charge density component of the current density might remain constant and the oscillation to actually affect the velocity. The resolution of this issue must include the treatment of the Josephson junction as a two level quantum system, which is subject to an external driving force and the law of conservation of charge. It leads to the observation that the time dependence of the charge densities ($\rho_j$ for $j=1,2$) in the two sides of the juction should be related by: $\dot{\rho}_1=-\dot{\rho}_2 \propto \sqrt{\rho_1 \rho_2} \sin{\phi},$ where $\phi$ is the phase difference of the condensate's wave function across the junction. The relation is a statement that if current flows from one of the sides, it charges up the other side which changes its electric potential. It is precisely this behaviour that leads to high frequency oscillations in the first place. If the junction is connected to a battery (constant electric potential $V$) currents will flow to keep the potential difference constant across the junction. When the currents from the battery are included in the Josephson junction description, standart solution dictates that charge densities do not change and there should be no mass oscillation\cite{S}. This is the famous $\rho_1= \rho_2={\rm const.}$ solution. However, this solution is correct provided $\sin{\phi}=0,$ which can be realised either by i.) $\phi \propto t$ and is {\it vanishing only as a mean value} (rapid oscillations which means that the average amplitude is assumed zero), ii.) $\phi = {\rm n} \pi$ for ${\rm n} \in\mathbb{Z},$ which is an exotic quantization condition, or iii.) $V=0,$ that is the static mode. Note, in general, the system of equations that describe the dynamics in the Josephson junction is highly non-linear. The standard solutions and description are mostly approximate. The DC mode is understood as $\rho_1= \rho_2={\rm const.}$ and $\phi=2eVt/\hbar,$ which leads to
\begin{eqnarray}
\dot{\rho}_1=-\dot{\rho}_2 \propto \sin{2eVt/\hbar} \neq 0
\end{eqnarray}
a contradiction due to the approximate character of the solution, but exactly the necessary behaviour for the emission of gravitational waves. Therefore,  charge (mass) density oscillation in a Josephson junction can take place in a stationary (DC mode), non-stationary (AC mode),  non-equilibrium electrical discharge or some form of Cooper pair destruction to automatically guarantee $\dot{\rho}_j \neq 0.$  It is a purpose of the paper to inspire the search for appropriate driving schemes which might be hidden in the non-linear sector of the Josephson junction dynamics as well.

The above feasibility calculation merely points to the possibility of channeling the Josephson energy using such a condensed matter system to generate gravitational wave emission (high frequency). Even though the superconducting Josephson junction is a quantum device, the behavior of the charged condensate is quasi-classical and the gravitational wave emission is therefore classical in its core. Condensed matter systems can in principle provide an environment in which high frequency mass oscillations (in the electronic subpart only) can occur and these oscillations can leverage the small mass that takes place in the conversion of kinetic energy into gravitational wave emission.

One last point concluding the discussion is the reverse effect, namely a large frequency gravitational wave can induce a charge oscillation in the same device which can be translated into DC voltage across the terminals, in effect executing the Josephson junction main property: frequency-to-voltage conversion. In this case the Josephson junction would extend its property to: curvature-to-voltage conversion.  Note, gravitational waves are not sensitive to electrical charges and would affect lattice ions and conducting electrons alike. As a result, the distortion of the superconducting material will be uniform and presumably undetectable, provided an extra mechanism is not at play. Such an additional mechanism is theoretically suggested to exist in Josephson junctions, where the curvature of space-time acts as a chemical potential and can create a potential difference for the superconducting species (not lattice ions) across the junction which translates as measurable voltage drop provided sufficiently large\cite{jj}. The reverse effect also represents a mean to detect the emission of gravitational waves with these high frequencies  and generated by a second sufficiently EMI insulated device at a distance.

The paper is intended to stimulate the experimental effort behind artificial gravitational wave induction and detection in condensed matter systems as a tool for studying gravitational field on a footing similar to the electromagnetic field. To put it in R.P. Feynman words: ''What I cannot create, I do not understand.''\cite{feynman} After the successful LIGO/VIRGO gravitational wave detection, the next logical step in gravitational exploration would be the attempt of artificial generation and detection (beyond the rotating frame) and the exploration of applications such as propulsion and communication.

\end{document}